\documentclass[journal]{IEEEtran}
\usepackage[hyphens]{url}
\usepackage{hyperref}
\hypersetup{breaklinks=true}
\hypersetup{draft}
\usepackage{gensymb}

\usepackage{comment}
\usepackage{enumitem}
\usepackage{multirow}
\usepackage{float}

\usepackage[margin = 0.8in]{geometry}
\usepackage{nopageno}
\usepackage{cite}
\usepackage{amsmath,amssymb,amsfonts}
\usepackage{algorithmic}
\usepackage{graphicx}
\usepackage{textcomp}
\usepackage[table]{xcolor}
\usepackage{listings}
\usepackage{color}

\ifCLASSINFOpdf
  \usepackage{graphicx}
  \graphicspath{{../Fig/}}
  \DeclareGraphicsExtensions{.pdf,.jpeg,.png}
\else
\fi
\usepackage{amsmath}
\newcommand\numberthis{\addtocounter{equation}{1}\tag{\theequation}}
\usepackage[linesnumbered,ruled]{algorithm2e}

\SetCommentSty{mycommfont}

\usepackage{caption}
\usepackage{array}
\usepackage{makecell}

\begin{document}
	\bstctlcite{IEEEexample:BSTcontrol}
	
	\raggedbottom
	
%
\title{Automotive Radar Interference Mitigation\\using Adaptive Noise Canceller}
%
%
%

\author{Feng~Jin,~\textit{Student Member, IEEE}~and~Siyang~Cao, \textit{Member, IEEE}
\thanks{Feng Jin and Siyang Cao are with Department of Electrical and Computer Engineering, \textit{the University of Arizona}, Tucson, AZ, 85719 USA. (e-mail: \{fengjin, caos\}@email.arizona.edu)}
}

%
%

\markboth{IEEE TRANSACTIONS ON VEHICULAR TECHNOLOGY, VOL. XX, NO. XX, XXX 2019}
{Shell \MakeLowercase{\textit{et al.}}: Bare Demo of IEEEtran.cls for IEEE Journals}
%



\maketitle

\begin{abstract}
Interference among frequency modulated continues wave (FMCW) automotive radars can either increase the noise floor, which occurs in the most cases, or generate a ghost target in rare situations. To address the increment of noise floor due to interference, we proposed a low calculation cost method using adaptive noise canceller to increase the signal-to-interference ratio (SIR). In a quadrature receiver, the interference in the positive half of frequency spectrum is correlated to that in the negative half of frequency spectrum, while the beat frequencies from real targets are always present in the positive frequency. Thus, we estimated the power of the negative frequency as an indication of interference, and fed the positive frequency and negative frequency components into the primary and reference channel of an adaptive noise canceller, respectively. The least mean square (LMS) algorithm was used to solve for the optimum filter solution. As a result, both the simulation and experiment showed a good interference mitigation performance.  
\end{abstract}


%
\IEEEpeerreviewmaketitle

\section{Introduction}
\IEEEPARstart{A}{s} people continue to seek safer and more comfortable driving experiences, the advanced driver-assistance systems (ADAS) and self-driving vehicle market is growing largely recently. The mmWave automotive radar is one of the key sensors in ADAS and self-driving system, due to its wide coverage, high reliability, all-weather and all-day capability of targets detection. As the frequency-modulated continuous wave (FMCW) is relatively easier to generate, by using voltage-controlled oscillator (VCO), and the frequency band of received echoes is narrow after the stretch processing, FMCW automotive radar is affordable to customers, and is widely accessed in the current market. Typically, each car is equipped with at least five automotive radar sensors for the ADAS functions, such as adaptive cruise control (ACC), blind spot detection (BSD) and cross traffic alert (CTA), to name a few. As the number of automotive radars on the road is increasing significantly, the interference among these radars will be more severe. 

\par The interference reduces the sensitivity of radar sensor or could potentially generate a ghost target, which in turn may cause missed or false detection. Hence, prior to 2012, an European funding project, MOSARIM (MOre Safety for All by Radar Interference Mitigation), had been conducted to investigate the possible interference mitigation methods for automotive radars \cite{ref_MOSARIM}. In this project, interference mitigation methods were classified into six domains as followings: (i) Polarization - In \cite{ref_PolarizedAntenna}, the transmitting antenna was designed in right hand circular polarized (RHCP) while the receiving antenna was in left hand circular polarized (LHCP). Thus, interference from the aggressor radar's transmitting antenna would be suppressed by the victim radar's receiving antenna. At the same time, the real target echo was received unhindered due to its polarization change to LHCP on account of the boundary conditions of electromagnetic fields on the surface of the target. (ii) Time domain - In \cite{ref_TimeDomainMethod,ref_TimeDomainMethod1,ref_TimeDomainMethod2}, the position of interference was found in the time domain, and a window-based method was used to remove the interference. In \cite{ref_Uhnder}, an adaptive filtering method in a phase modulated continuous wave (PMCW) radar system was proposed to mitigate interference from FMCW radar. In \cite{ref_MCA}, the interference was removed by using morphological component analysis (MCA). (iii) Frequency domain - In \cite{ref_BATS}, the interference was avoided by changing the transmitting signal's frequency band after detecting the interference's frequency band. In \cite{ref_InterpolationSTFT}, the signal-to-noise ratio (SNR) was improved by interpolating the beat frequency in the short time Fourier transform (STFT) domain. (iv) Coding techniques. In \cite{ref_CODING,ref_CODING1,ref_CODING3}, the orthogonality property of random coded chirp sequence leads to inherent interference immunity. (v) Space domain - In \cite{ref_DBF,ref_DBF1,ref_DBF2}, the adaptive beamforming method was used to mitigate the interference from side lobes. (vi) Strategic approach - In \cite{ref_MAC}, a control center was setup to receive location/speed information from all the radars, and dispatch waveform parameters to each one to avoid the interference among them.

\par As we can see, those methods which depend on complicated antenna design, more antenna channels, higher ADC sampling rate and a control center are not preferred to the cost sensitive automotive market. Therefore, we are proposing a new interference mitigation method using adaptive filtering on the frequency domain, which works on the current and mostly used FMCW radar sensors and does not require costly hardware improvements. 

\par The adaptive filtering method is widely used to retrieve the target signals from additive interference since it was first introduced by B. Widrow \cite{ref_WIDROW}. Currently, the adaptive filtering on FMCW radar interference mitigation draws little attention to the researchers and engineers. In Uhnder's patent \cite{ref_Uhnder}, the adaptive line enhancer was introduced in their PMCW radar system to mitigate the interference from FMCW radars. By using advanced line enhancer, the priori assumption is that the interference is close to Gaussian white noise, which means its delayed version is not correlated to itself. This may be true for the interference caused by FMCW radar in the PMCW radar system. However, the interference among FMCW radars, which is the most common case currently, does not adherent to this property. Hence, the advanced line enhancer cannot be applied to suppress interference among FMCW radars.

\par In this paper, we introduce the adaptive noise canceller method to deal with the interference among FMCW automotive radars. In the quadrature receiver, the interference in the positive half of frequency spectrum ([0, $\pi$] normalized frequency) is correlated to that in the negative half of frequency spectrum ([$-\pi$, 0] normalized frequency), while the real target echo is only present in the positive half of frequency spectrum. After applying range-FFT (fast Fourier transform on the fast time data), the positive half of FFT ([0, $N$/2-1], $N$ is FFT size) and the negative half of FFT ([$N$/2, $N$-1], $N$ is FFT size) are separated first. In addition, the power of the negative half of FFT is estimated as an indication of interference. Then the positive and negative half of FFT are fed into the primary and reference channel of the adaptive noise canceller, respectively. And the least mean square (LMS) method is used to find the optimum filter solution. Finally, this study found that the interference could be suppressed significantly.

\par The organization of this paper is as follows. Section II summarizes the background of FMCW automotive radar and the adaptive filtering. Two types of interference among FMCW automotive radars are discussed in section III. In Section IV, the interference mitigation method using adaptive noise canceller is proposed. Both the $Matlab$ simulation and field experiment are conducted and discussed in section V. Section VI concludes this paper.

\section{Background of FMCW Automotive Radar and Adaptive Noise Canceller}
\subsection{Concepts of FMCW Automotive Radar with Quadrature Receiver}
\par
\begin{figure}[h]
	\centering
	\scalebox{0.5}{
	\includegraphics{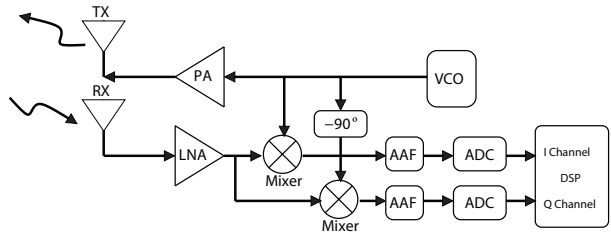}}
	\caption{Architecture of FMCW automotive radar with quadrature receiver.}
	\label{Figure1}
\end{figure}

 Prior to early 1970s, several European companies had started to investigate radar technology for vehicles \cite{ref_TRENDS1998}. However, these discrete waveguide components were pricey and bulky and made it unsuitable for automotive applications. After that, the industry tried to develop the radar sensor into monolithic microwave integrated circuit (MMIC) based on GaAs \cite{ref_MMIC} and SiGe \cite{ref_MMIC_SIGE} process to shrink the size and price of the components. In 2008, Infineon introduced its first fully integrated MMIC which integrated the VCO, frequency multiplier, power amplifier (PA), low noise amplifier (LNA) and mixer together \cite{ref_MMIC_Infineon}. In 2017, Texas Instruments (TI) introduced the most integrated automotive radar MMIC so far, which integrates not only the size effective mmWave parts but also a signal processing chain, like anti-aliasing filter (AAF), ADCs, DSP and ARM processor, into a single chip \cite{ref_MMIC_TI}. At the same time the quadrature mixer and complex baseband were implemented in such a MMIC, which has more signal processing benefits than the traditional single channel receiver \cite{ref_TIPaper}. Fig. \ref{Figure1} shows the architecture of a typical FMCW automotive radar with quadrature receiver. 

\par
\begin{figure}[h]
	\centering
	\scalebox{0.5}{
	\includegraphics{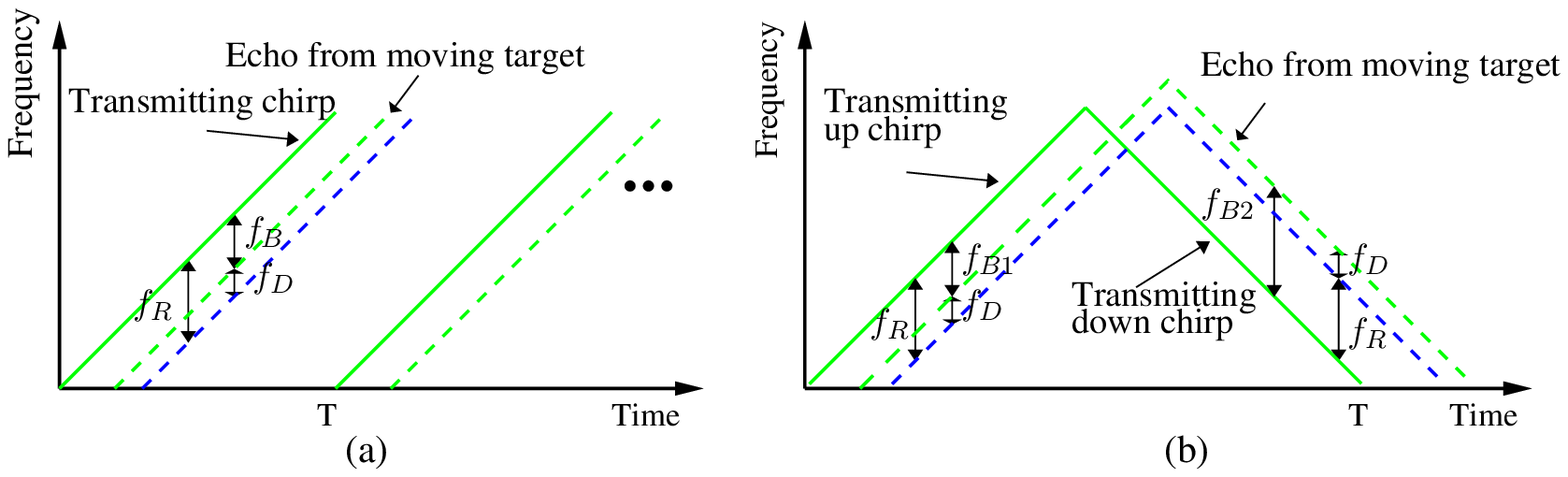}}
	\vspace*{-0.3cm}
	\caption{FMCW waveforms, \(f_R\) is the frequency difference due to the range, \(f_D\) is the frequency difference due to the Doppler effect, and \(f_B\)/\(f_{B1}\)/\(f_{B2}\) is the beat frequency after stretching processing. (a) Chirp sequence FMCW. (b) Triangular FMCW.}
	\label{FMCW}
\end{figure}

 Due to the easy generation and simple processing, two types of FMCW waveform are widely used in the current automotive radar sensor, as shown in Fig. \ref{FMCW}. After stretching processing, the beat frequency with range-Doppler coupling is obtained. Chirp sequence FMCW uses coherent Doppler processing across all the chirps during one coherent processing interval (CPI) to solve for the Doppler frequency \(f_D\). For the triangular FMCW, as the beat frequency during up chirp is \(f_{B1}=f_R-f_D\) and the beat frequency during down chirp is \(f_{B2}=f_R+f_D\), range and Doppler coupling can be resolved. More variants of FMCW waveform can be found in \cite{ref_FMCWWaveform1}, \cite{ref_FMCWWaveform2}. 

\subsection{LMS based Adaptive Noise Canceller}\label{2b2}
\begin{figure}[h]
	\centering
	\scalebox{0.6}{
		\includegraphics{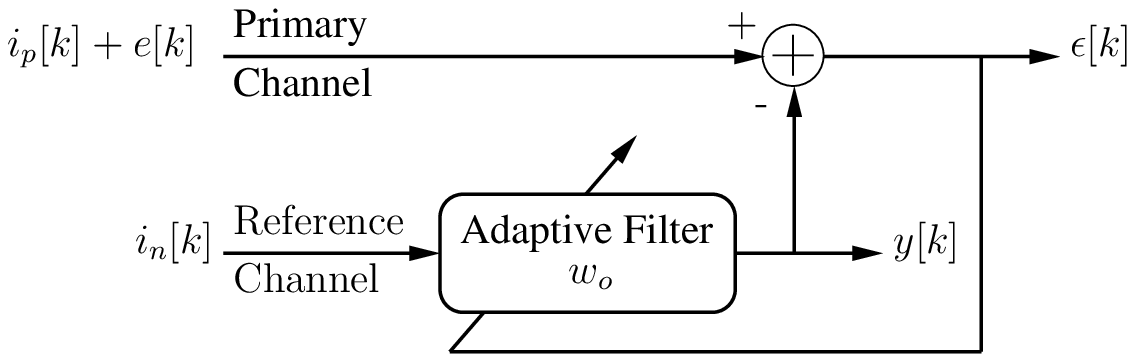}}
	\caption{Architecture of the adaptive noise canceller.}
	\label{Figure3}
\end{figure}

\par The adaptive noise canceller was well studied and discussed in detail in \cite{widrow1985adaptive}. As in Fig. \ref{Figure3}, when the adaptive filter outputs an optimum estimation of \({i_p}[k]\) from a reference input \({i_n}[k]\), the \({i_p}[k]\) can be cancelled from the primary channel, which makes $\epsilon[k]$ become an optimum estimation of $e[k]$. Here, \({i_p}[k]\) and \({i_n}[k]\) are jointly wide-sense stationary (WSS) stochastic processes with zero mean and uncorrelated to \(e[k]\). 

\par More concretely, at instant \(k\),
\begin{equation}
\epsilon=e+i_p-y=e+i_p-\mathbf{w_o}^T\mathbf{i_n}(k).
\label{equ_ANCEstimationError}
\end{equation}
where \(\mathbf{i_n}(k)\) denotes \(L\)-by-\(1\) input vector of the adaptive filter, and $y[k]$ denotes the output, \(L\) is the filter length and \(\mathbf{w_o}\) is the \(L\)-by-\(1\) tape-coefficient vector.

\par By applying statistical expectation on both sides, we have
\begin{equation}
\begin{aligned}
&E[\epsilon^2]=E[(e+i_p-y)^2]\\
&=E[e^2]+E[(i_p-y)^2+2E[e*i_p]-2E[e*y]\\
&=E[e^2]+E[(i_p-y)^2+2E[e]E[i_p]-2E[e]E[y]\\
&=E[e^2]+E[(i_p-y)^2]\\
&=E[e^2]+E\{[i_p-\mathbf{w_o}^T\mathbf{i_n}(k)]^2\}.
\end{aligned}
\label{equ_ErrorExpectation}
\end{equation}

\par Furthermore,
\begin{equation}
\begin{aligned}
min\{E[\epsilon^2]\}&=min\{E[e^2]+E[(i_p-y)^2]\}\\
&=E[e^2]+min\{E[(i_p-y)^2]\}.
\end{aligned}
\label{equ_minErrorExpectation}
\end{equation}
\par According to Eq. (\ref{equ_ANCEstimationError}),
\begin{equation}
\epsilon-e=i_p-y.
\end{equation}
\par Thus,
\begin{equation}
min\{E[(\epsilon-e)^2]\}=min\{E[(i_p-y)^2]\}.
\label{equ_minErrorPrimaryInf}
\end{equation}

\par According to Eq. (\ref{equ_minErrorExpectation}) and (\ref{equ_minErrorPrimaryInf}), when \(min\{E[\epsilon^2]\}\) is achieved, \(min\{E[(\epsilon-e)^2\}\) is achieved as well, which means \(\epsilon[k]\) is an estimation of \(e[k]\) in a minimum mean square error (MMSE) sense. As a result, the goal is to find a optimum solution $\mathbf{w_o}$ to minimize \({E[\epsilon^2]}\).

\par The LMS method \cite{ref_HAYKIN} is a practical way to solve for the minimum of \(E[\epsilon^2]\) by moving the filter taps $\mathbf{w}$ to $\mathbf{w_o}$ with a small step $\Delta w$ iteratively. Finally, the adaptive noise canceller with LMS algorithm could be summarized as followings:
\begin{align*}
Filter~Output:&\hspace{0.1cm} y[k]=\mathbf{w}^T[k]\mathbf{i_n}[k], \numberthis \\
Est.~Error:&\hspace{0.1cm} \epsilon[k]=e[k]+i_p[k]-y[k], \numberthis \\
Tap~Updates:&\hspace{0.1cm} \mathbf{w}(k+1)=\mathbf{w}(k)+\Delta w\mathbf{i_n}[k]\mathbf{\epsilon}^*[k]. \numberthis
\end{align*}
where \(\Delta w\) controls the convergence speed and stability of the adaptive filter. To have a convergence result, \(\Delta w\) should satisfy
\begin{equation}
0<\Delta w<\frac{2}{input\ power}=\frac{2}{\sum_{l=0}^{L-1}E\{|i_n[k-l]|^2\}}.
\label{equ_ConvergencePower}
\end{equation}

\section{Mutual Interference between FMCW Automotive Radars with Quadrature Receiver}
\begin{figure}[h]
	\centering
	\scalebox{0.5}{
	\includegraphics{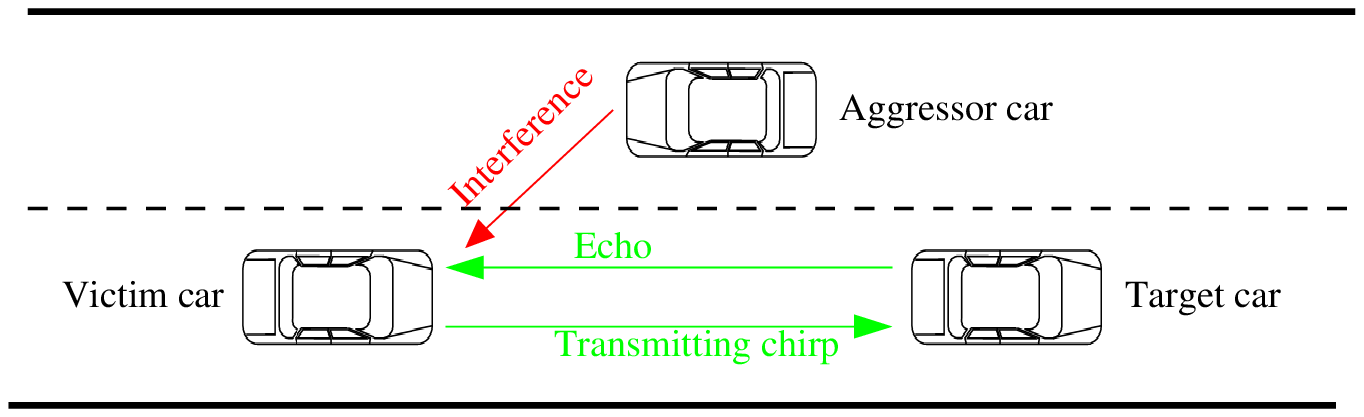}}
	\caption{Interference in a road scenario.}
	\label{Figure4}
\end{figure}

Mutual interference occurs when at least two vehicles confront each other in the road scenario like Fig. \ref{Figure4}. The interference can be generally classified into two categories based on whether the waveform used in the aggressor radar is different from the waveform used in the victim radar. 
\subsection{Two FMCW Automotive Radars with Different Parameters}\label{3b1}
\begin{figure}[h]
	\centering
	\scalebox{0.5}{
	\includegraphics{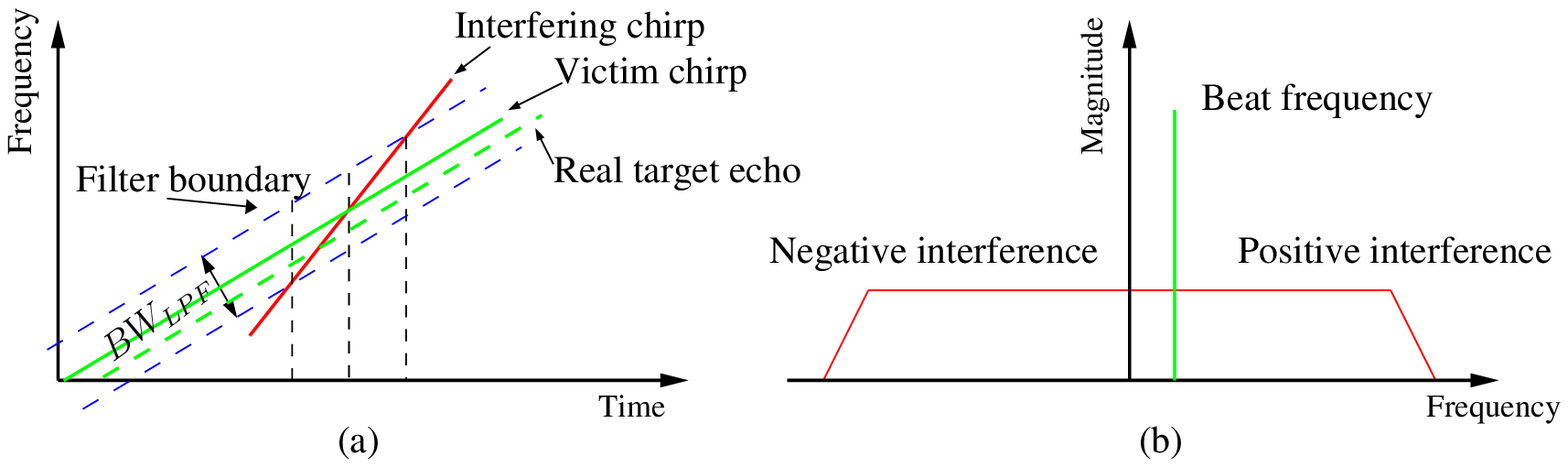}}
	\vspace*{-0.3cm}
	\caption{Interference scenario between two different FMCW automotive radars. (a) Time-frequency plot. (b) Frequency domain.}
	\label{Figure5}
\end{figure}

This is the most common case, considering the interference between the triangular FMCW and chirp sequence FMCW, staggered PRFs used to solve for the velocity ambiguity, long range or short range applications, etc. In this scenario, the received interference by victim radar is filtered by the AAF, mostly low pass filter (LPF), although some real applications might require band pass filter (BPF) to eliminate close range clutter to avoid ADC saturation. This results in a linear frequency modulated (LFM) like interference as shown in Fig. \ref{Figure5}. When the frequency of victim chirp is above the interfering chirp, it generates the interference in the positive half of frequency. Otherwise, a negative interference appears. On the other hand, because the real target echo is always a time-delayed version of the transmitting chirp, the beat frequency is always in the positive half of frequency. As a result, the signal-to-interference (SIR) is decreased, and the sensitivity of radar sensor is reduced.  
\par Furthermore, considering the FFT data from the time series point of view, the beat frequency is a non-stationary process because it is always a single tone, which means it changes sharply over FFT bins. On the other hand, the interference appears as a WSS process due to its LFM-like shape changing slowly over FFT bins. The chirp has a zero mean in the time domain, and thus the FFT of interference are also zero mean, according to Eq. (\ref{equ_FFTZeroMean}). As a result, the FFT of interference can be assumed as a zero mean WSS process. And because the interference in the positive and negative half of FFT results from the same interfering source, it is reasonable to suggest that they are correlated. At the same time, the analog AAF is real which means its frequency response is symmetric, thus ideally the interference in the negative half of FFT is conjugate symmetric to that in the positive half of FFT. 
\begin{figure}[h]
	\centering
	\includegraphics[width=3in]{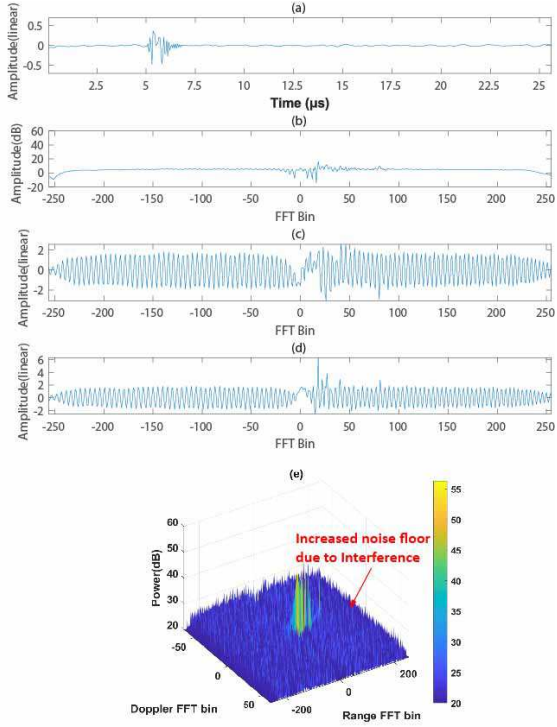}
	\caption{Experimental data of interference between two different FMCWs. (a) Raw data (only real part) in the time domain. (b) The magnitude of FFT of the raw data. (c) The real part of FFT. (d) The image part of FFT. (e) Range-Doppler plot.}
	\label{Figure6}
\end{figure}
\begin{equation}
\begin{aligned}
E\{X[k]\}&=E\{\sum_{n=0}^{N-1}x[n]e^{-j\frac{2\pi}{N}kn}\}\\
&=\sum_{n=0}^{N-1}E\{x[n]\}e^{-j\frac{2\pi}{N}kn}=\sum_{n=0}^{N-1}0*e^{-j\frac{2\pi}{N}kn}=0.
\end{aligned}
\label{equ_FFTZeroMean}
\end{equation}
\par To summarize, we have several assumptions, as the followings:  
\begin{enumerate}[label=(\roman*)]
	\item The FFT of echoes from multiple targets is an non-stationary complex random process.
	\item The FFT of interference from multiple interfering sources is a zero-mean WSS complex random process.
	\item The FFT of echoes and the FFT of the interference are uncorrelated.
	\item The interference in negative and positive half of FFT are correlated. Ideally, they are conjugate symmetric.
\end{enumerate}
\par We conducted an experiment to collect radar data, in order to verify the assumptions above. Extensive details of the experiment has been present in Sec. \ref{5b2}. In this experiment, the victim radar was sweeping from 77~GHz to 77.75~GHz over one chirp of 29.56 $\mu s$, and the aggressor radar was sweeping from 77~GHz to 77.682~GHz over 72.31 $\mu s$. The captured interference is shown in Fig. \ref{Figure6}. From the spectrum plot in Fig. \ref{Figure6} (b), the interference in the negative and positive half of FFT exhibit the WSS property, and appear symmetric to each other. Fig. \ref{Figure6} (c) and (d) show the zero mean property appearance.   
\begin{figure}[h]
	\centering
	\includegraphics[width=2.9in]{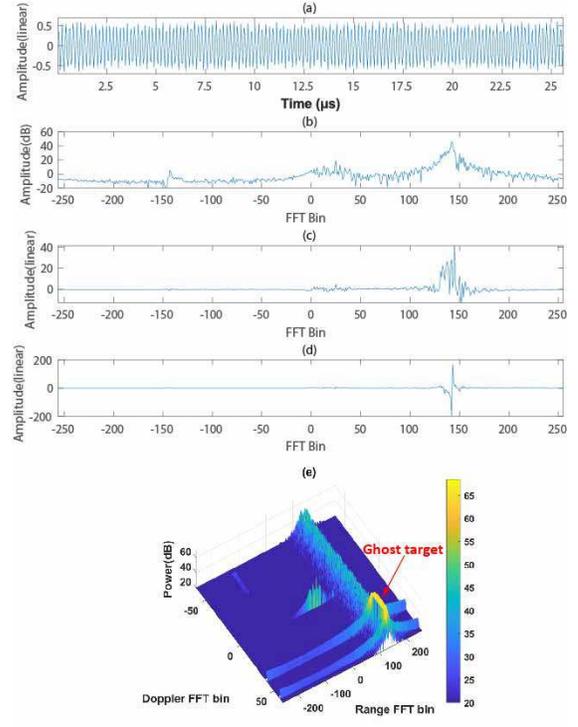}
	\caption{Experimental data of ghost target scenario. (a) Raw data (only real part) in the time domain. (b) The magnitude of FFT of the raw data. (c) The real part of FFT. (d) The image part of FFT. (e) Range-Doppler plot.}
	\label{Figure7}
\end{figure}
\subsection{Two FMCW Automotive Radars with Identical Parameters} 
\begin{figure}[h]
	\centering
	\scalebox{0.5}{
		\includegraphics{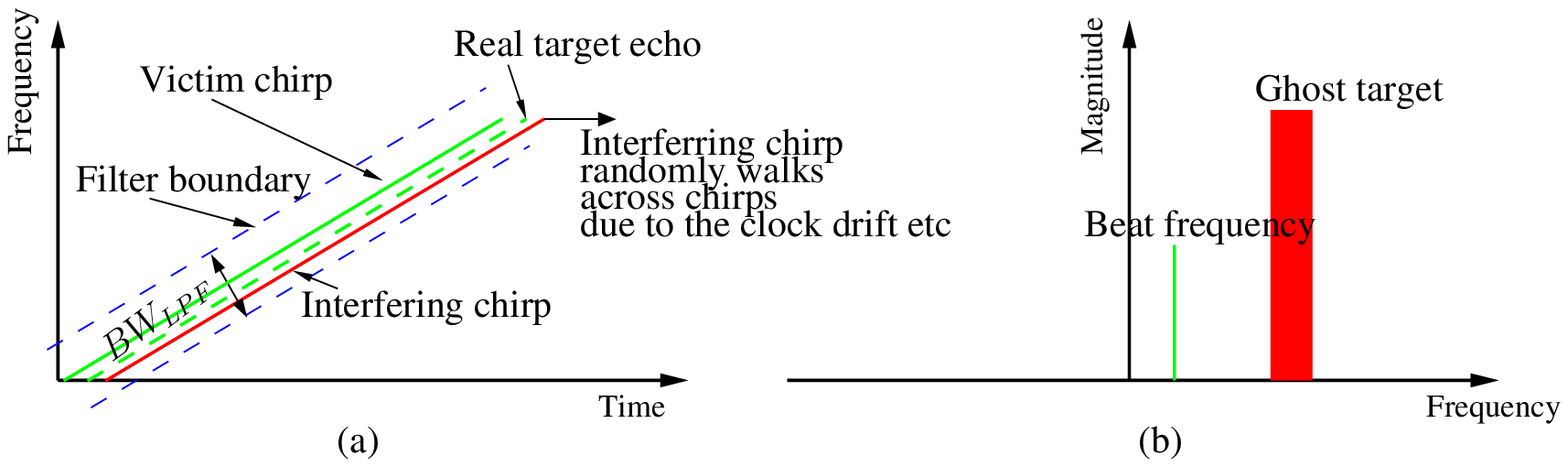}}
	\vspace*{-0.3cm}
	\caption{Interference scenario of ghost target. (a) Time-frequency plot. (b) Frequency domain.}
	\label{Figure8}
\end{figure}
As shown in Fig. \ref{Figure8}, due to the random phase relationship between the two radars, the ghost target occurs only when the frequency difference between the interfering chirp and victim chirp falls into AAF's bandwidth. Thus the occurring probability of ghost target can be expressed as:
\begin{equation}
P_{GhostTarget}=\frac{BW_{LPF}}{BW}.
\label{equ_PGhostTarget}
\end{equation}
\par For the waveform configuration used in the experiment, the receiver bandwidth $BW_{LPF} = 9$~MHz and sweeping bandwidth $BW$=$750$~MHz, then $P_{GhostTarget}$=$1.2\%$. The probability can be even less for low range (low \(BW_{LPF}\)) high resolution (high \(BW\)) application. Other researchers gave the ghost target probability of less than 0.000665 in \cite{ref_BrookerInterference} and of \(10^{-3}\) in \cite{ref_RohlingInterference}. To resolve the velocity ambiguity, staggered PRFs would be used, which leads to lower possibility of two same chirps interfering each other in real case.  
\par In practice, due to the internal clock drifting and carrier frequency offset \cite{ref_RohlingInterference}, the phase between the interfering chirp and victim chirp is randomly walking during one CPI. This leads to the ghost target moves across range bins and has a broad Doppler shift, as the target peak is expanded in both range and Doppler dimension in the frequency domain. Fig. \ref{Figure7} shows the experimental data in which both the victim and aggressor radar were sweeping from 77~GHz to 77.75~GHz over 29.56 $\mu s$. In this case, we do not see the interference exhibiting WSS, symmetric appearance and zero mean.

\section{Proposed Interference Mitigation Method}
\par
\begin{figure}[h]
	\centering
	\scalebox{0.5}{
	\includegraphics{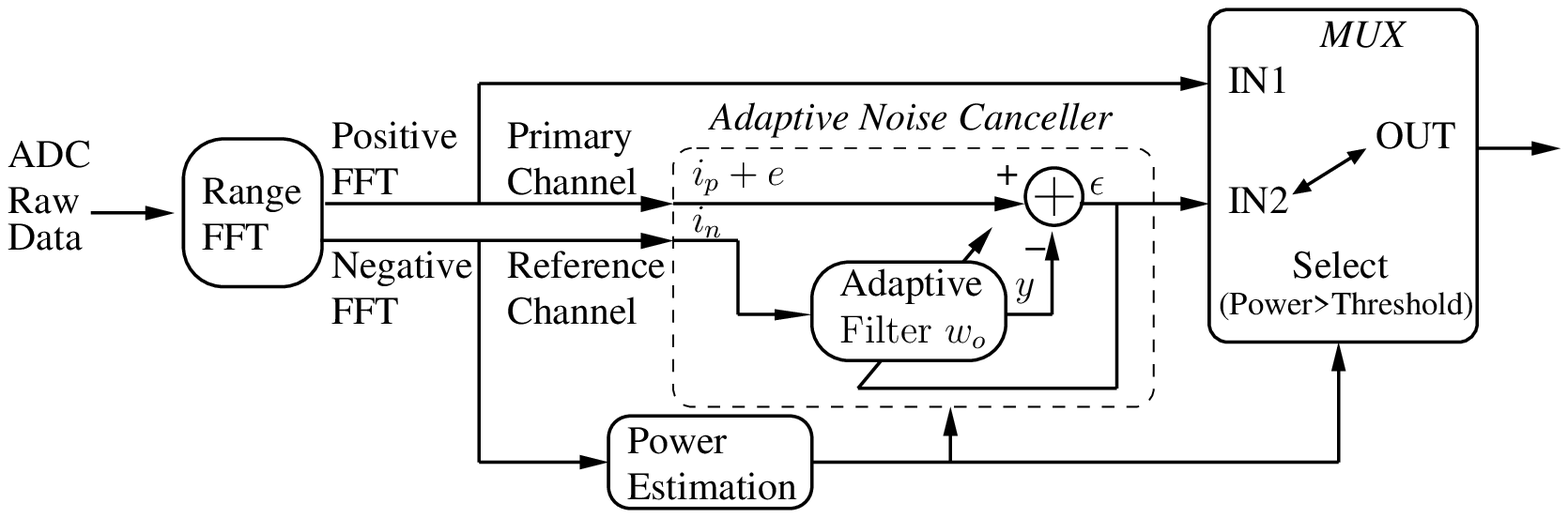}}
	\caption{Proposed interference mitigation scheme using adaptive noise canceller.}
	\label{Figure9}
\end{figure}

 To deal with the interference represented in Fig. \ref{Figure5}, which is also the most commonly occurred scenario, we now propose a novel adaptive noise canceller (ANC) based interference mitigation method by cancelling the interference in the positive half of frequency using the correlated interference in the negative half of frequency as a reference. The proposed system is shown in Fig. \ref{Figure9}. 

\par As in Fig. \ref{Figure9}, \(e[k]\) denotes the beat frequencies due to the real target echo, \(i_p[k]\) and \(i_n[k]\) denotes the interference in the positive and negative half of FFT respectively, where $k$ is FFT bin. According to the discussion and assumptions in Sec. \ref{3b1}, \(e[k]\) is a non-stationary process and uncorrelated to the zero-mean WSS processes \(i_p[k]\) and \(i_n[k]\). Meanwhile, \(i_p[k]\) and \(i_n[k]\) are correlated, and ideally they are conjugate symmetric. Thus, the LMS based ANC method discussed in Sec. \ref{2b2} can be used to mitigate the interference \(i_p[k]\) in the positive half of FFT.

\par After range-FFT processing (fast time FFT), the \textit{Power} of \(i_n[k]\) is estimated to identify the existence of interference, when the estimated \textit{Power} is greater than a \textit{Threshold}. If there is no interference, the ANC will be bypassed to save computational resources. Otherwise, the ANC is activated, and the estimated \textit{Power} is used to update the adaptive filtering step size \(\Delta w\) according to Eq. (\ref{equ_ConvergencePower}). Finally, Algorithm \ref{alg1} lists the detailed computing procedure.
\begin{algorithm}
	\caption{Interference Mitigation using Adaptive Noise Canceller (ANC)}
	\label{alg1}
	\KwIn{$\textbf{IN}$, radar raw data ($M$$\times$$N$, $N$ fast time samples per chirp, $M$ number of chirps per CPI). ${T}$, interference power threshold. $L$, filter length. $\gamma$, fraction of the upper bond of step size $\Delta w$.}
	\KwOut{$\textbf{OUT}$ $M\times(N/2)$ filtered range-FFT data}
	\BlankLine
	\For{$i=1$ to $M$} {
		$y$ = FFT($\textbf{IN}(i, :)$, $N$); \tcp*[h]{Apply range-FFT}\\
		$pri$ = $y$(0:$N$/2-1); \tcp*[h]{Positive FFT} \\
		\tcp*[h]{Conjugate symmetry of negative FFT} \\
		$ref$ = conjugate(flip($y$($N$/2:$N$-1))); \\
		\tcp*[h]{Interference power} \\
		$P=\sum_{n=0}^{N/2-1}\{|ref(n)|^2\}$; \\
		\eIf(\tcp*[h]{Apply ANC}){$P > T$}{
			$\mathbf{w_o}=(1,0,...,0)^T$; \tcp*[h]{Filter taps $L\times1$}\\
			$\mathbf{f_i}=(0,0,...,0)^T$; \tcp*[h]{Filter input $N\times1$}\\
			$\mathbf{\epsilon}=(0,...,0)^T$; \tcp*[h]{Est. error $N\times1$}\\
			$\Delta w = \frac{2}{\gamma*P}$; \tcp*[h]{Set Step size}\\
			\For{$j=1$ to $N$} {
				$\mathbf{f_i}$=[$ref(j)$\ $\mathbf{f_i}$(1:$L$-1)]; \tcp*[h]{Filter input}\\	
				$f_o=\mathbf{w_o}^T*\mathbf{f_i}$; \tcp*[h]{Filter output}\\	
				$\epsilon(j)=pri(j)-f_o$; \tcp*[h]{Est. error}\\
				$\mathbf{w_o}=\mathbf{w_o}+\Delta w*\mathbf{f_i}*\epsilon^*$; \tcp*[h]{Tap update}\\
			}
			$\textbf{OUT}(i, :)$ = $\epsilon$;
		}
		(\tcp*[h]{Bypass ANC}){
			$\textbf{OUT}(i, :)$ = $pri$;
		}
	
	}
\end{algorithm}

\section{Simulation and Experiment Result of the Proposed Method}
In this section, we conducted (i) a simulation on $Matlab$ and (ii) a field experiment with a TI mmWave board AWR1243BOOST, to investigate the effectiveness of the proposed mitigation method for the interference among FMCW automotive radars.
\subsection{$Matlab$ Simulation}
\par
\begin{figure}[h]
	\centering
	\includegraphics[width=3in]{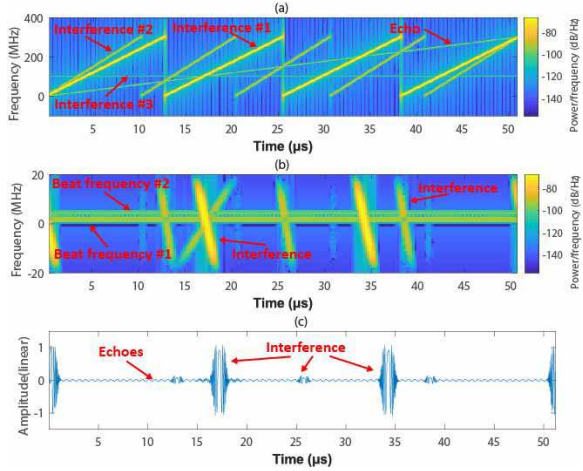}
	\caption{The simulated received signal. (a) The STFT of received signal before the mixer. (b) The STFT of received signal after the mixer. (c) The time domain of received signal (only real part).}
	\label{STFT}
\end{figure}

 In this simulation, the victim radar was configured as a typical long range radar (LRR) with the starting frequency of 76~GHz \cite{ref_SAE} and 300~MHz sweeping bandwidth of 0.5~meters range resolution ($\Delta R$~=~$\frac{c}{2BW}$, $c$ is the speed of light). The receiver LNA gain was set to 40~dB, a linear-phase FIR LPF was emulated to the AAF with the passband frequency of 10~MHz and the stopband frequency of 20~MHz, which leads to 256~meters unambiguous range ($\frac{2R_{un}}{c}\mu$~$\leq$~$BW_{LPF}$). After that, the signal was subjected to a 40~MHz ADC. The static target \#1 was set at 40~meters, and the target \#2 with distance of 100~meters had three times normalized RCS of target \#1's for better demonstration, in case that its SIR was too low to be visible. There were three different sources of interference as listed in Table \ref{tab_SimEchoInf}. As seen from the short time Fourier transform (STFT) of received signal in Fig. \ref{STFT}, the beat frequencies were significantly impaired by the interference. And the interference appears like LFM signals, when plotted in the time domain. 
\begin{table}[H]
	\caption{Simulation Configuration. $f_c$, start frequency. $BW$, sweeping bandwidth. $T$, chirp duration. $\mu$, chirp rate. $f_s$, ADC sampling rate. $BW_{LPF}$, receiver bandwidth. $N$, fast time FFT size. $d$, distance. $\sigma$, normalized RCS.}
	\label{tab_SimEchoInf}
	\centering
	\scalebox{0.85}{\begin{tabular}{|c|c|c|c|c|c|c|c|}
		\hline
		\bfseries Item &  \bfseries Para &  \bfseries Value &  \bfseries Unit & \bfseries Item &  \bfseries Para &  \bfseries Value &  \bfseries Unit \\ \hline
		\multirow{11}{*}{} & $f_{c}$ & 76 & GHz  & &  $f_{c}$ & 76 & GHz  \\ \cline{2-4} \cline{6-8} 
		& $BW$ & 300 &  MHz  & Inf1 &  $T$ & 10 &  us  \\ \cline{2-4} \cline{6-8} 
		& $T$ &  51.2 & us  & & $\mu$ &  30 & MHz/us   \\ \cline{2-4} \cline{6-8} 
		\makecell{FM-\\CW} &  $\mu$ &  5.86 & MHz/us & & $d$ &  10 & meter 	\\ \cline{2-8} 
		& $f_s$ &  40 & MHz  &	&  $f_{c}$ & 76 & GHz    \\ \cline{2-4}  \cline{6-8}
		& $BW_{LPF}$ &  10 & MHz & Inf2 &  $T$ & 8 &  us  \\ \cline{2-4}  \cline{6-8}
		& $N$ &  2048 &  &  & $\mu$ &  37.5 & MHz/us   \\ \cline{1-4} \cline{6-8}
		&  $d$ & 35 &  meter &  & $d$ &  20 & meter  \\ \cline{2-4} \cline{5-8}
		T1 &  $\sigma$ & 1 &  dBsm & &  $f_{c}$ & 76.1 & GHz \\ \cline{1-4} \cline{6-8}
		&  $d$ & 100 &  meter &  Inf3  & $\mu$ &  0 & MHz/us \\ \cline{2-4} \cline{6-8}
		T2 &  $\sigma$ & 4 &  dBsm &  & $d$ &  30 & meter \\ \hline 
	\end{tabular}}
\end{table}
\begin{figure}[h]
	\centering
	\includegraphics[width=3in]{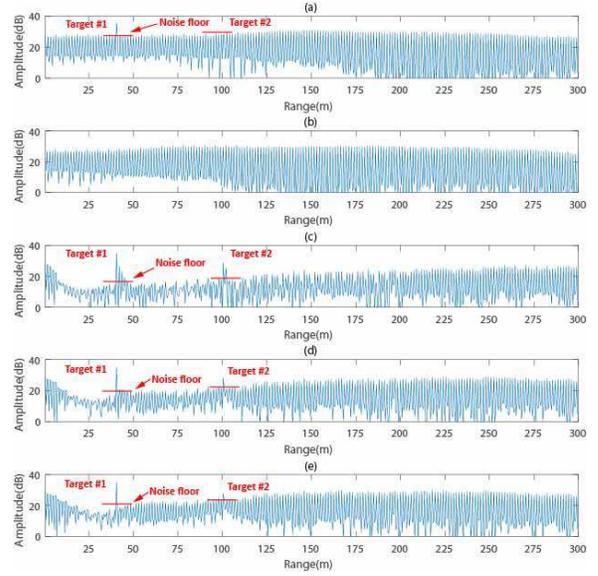}
	\caption{The simulation using the proposed method for different adaptive filtering step size $\Delta w$. (a) The positive half of FFT of received signal. SIR1~=~12.42~dB, SIR2~=~3.71~dB. (b) The conjugate symmetry of negative FFT of received signal. (c) The result with $\Delta w$~=~$\frac{2}{50*P}$. SIR1~=~20.40~dB, SIR2~=~14.40~dB. (d) The result with $\Delta w$~=~$\frac{2}{100*P}$. SIR1~=~19.31~dB, SIR2~=~9.89~dB. (e) The result with $\Delta w$~=~$\frac{2}{150*P}$. SIR1~=~19.14~dB, SIR2~=~7.13~dB.}
	\label{Figure11}
\end{figure}

\par In Fig. \ref{Figure11}, the target \#1 was visible, however the target \#2 was completely buried in the interference. Meanwhile, there was no beat frequencies, just interference in the negative FFT. The estimated negative interference power was about $P$~=~$22$~dB. A typical constant false alarm rate (CFAR) window with 20 reference cell and 6 guard cells was referred to calculate the SIR. With a large $\Delta w$ as seen in Fig. \ref{Figure11} (c), both the two targets' SIR increased significantly compared to the original one, as shown in Fig. \ref{Figure11} (a). However, there were some side lobes alongside the target, which may result in false target detection. This is because of a large step size $\Delta w$, that resulted in the final filter solution to deviate from the optimum Wiener filter after finite number of iterations. When $\Delta w$ was decreased as shown in Fig. \ref{Figure11} (c), the side lobes disappeared, albeit at a lesser SIR improvement. Continuing to decrease as shown in Fig. \ref{Figure11} (c), the SIR improvement dropped furthermore such that the target \#2 was almost lost for detection. Therefore, with an ideal medium step size $\Delta w$=$\frac{2}{100*P}$, the side lobe effect was eliminated, and the SIR of target \#1 and \#2 was increased by 6.89~dB and 6.18~dB, respectively.

\subsection{Field Experiment}\label{5b2}
\begin{figure}[H]
	\centering
	\includegraphics[width=2.5in]{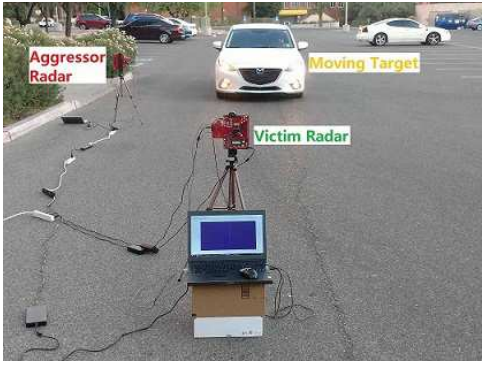}
	\caption{Field experiment setup.}
	\label{Figure12}
\end{figure}

In the field experiment, we adopted the TI AWR1243BOOST mmWave radar development board \cite{ref_TIBOOST} for data capture. Fig. \ref{Figure12} shows the experiment setup in a parking lot. The victim radar was configured as a typical short range radar (SRR) with a maximum unambiguous range of 46~meters and range resolution of 0.2~meters. One white car was driving back and forth to act as a moving target. And the aggressor radar with a \(\frac{1}{3}\) chirp rate of that of the victim radar was set as the interfering source. The radar configuration set for the experiments are listed in Table \ref{tab_ExperimentWaveform}. Note that the sweeping bandwidth $BW$ was calculated by $\frac{N}{f_s}\mu$, not $T\mu$. This is to account for the delay in ADC capture starting time with respect to the chirp beginning time. To have a significant interference power, we put the aggressor radar in front of the victim radar at a distance of 2 meters. The captured data was saved to disk and processed on $Matlab$.
\begin{table}[H]
	\caption{Experiment Configuration. $f_c$, start frequency. $BW$, sweeping bandwidth. $T$, chirp duration. $\mu$, chirp rate. $f_s$, ADC sampling rate. $BW_{LPF}$, receiver bandwidth. $N$, number of fast time samples. $M$, number of chirps per CPI.}
	\label{tab_ExperimentWaveform}
	\centering
	\scalebox{0.85}{\begin{tabular}{|c||c||c||c|}
		\hline
		\bfseries Parameter & \bfseries Victim Radar & \bfseries Aggressor Radar & \bfseries Unit \\
		\hline
		$f_c$ & 77 & 77 & GHz\\
		\hline
		$BW$ & 750 & 682 & MHz\\
		\hline
		$T$ & 29.56 & 72.31 & us\\
		\hline
		$\mu$ & 29.306 & 9.994 & MHz/us\\
		\hline
		$f_s$ & 20 & 15 & MHz\\
		\hline
		$BW_{LPF}$ & 9 & 6.75 & MHz\\
		\hline
		$N$ & 512 & 1024 &\\
		\hline
		$M$ & 128 & 128 &\\
		\hline
	\end{tabular}}
\end{table}
\begin{figure}[H]
	\centering
	\includegraphics[width=3.5in]{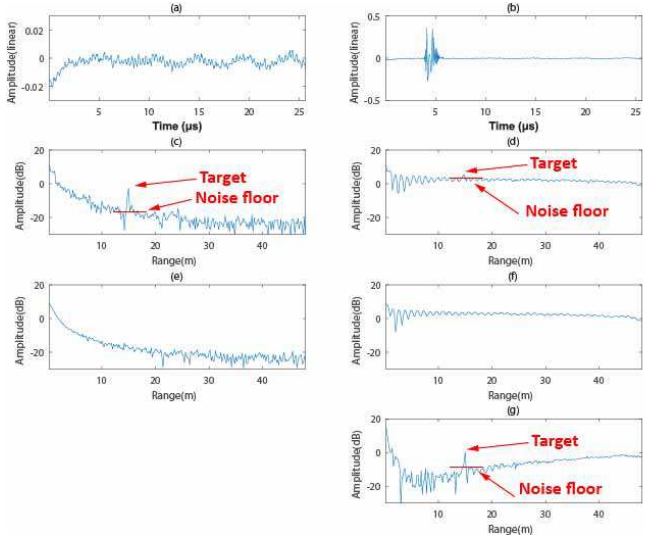}
	\caption{Data process of one chirp in the field experiment. (a) The 10th chirp (without interference, only real part) in the time domain. (b) The 50th chirp (with interference,  only real part) in the time domain. (c) The positive half of FFT of 10th chirp, SIR=10.52~dB. (d) The positive half of FFT of 50th chirp, SIR=2.95~dB. (e) The conjugate symmetry of negative FFT of the 10th chirp. (f) The conjugate symmetry of negative FFT of the 50th chirp. (g) The output of proposed method on the 50th chirp, SIR=10.55~dB.}
	\label{Figure13}
\end{figure}

\par With the collected experiment data, the proposed method was first applied on one chirp to check the effectiveness of the proposed methodology. The results are shown in Fig. \ref{Figure13}. There were 128 chirps per CPI in the experiment, however we observed that the interference was not present in all the chirps in a CPI. This is due to the fact that the phase relationships between these two radars may cause the interference to be out of the victim radar's receiver passband. As we can see in Fig. \ref{Figure13} (a), there was no interference, but only echoes and static clutter returns in the 10th chirp. However, the interference was observed to be present in the 50th chirp. As a result, in the positive half of FFT plot, the SIR of target at about 15 meters in the 10th chirp was much higher than the same target in the 50th chirp. Meanwhile, there was no target but only noise and/or interference in the negative half of FFT of both chirps. Upon studying the negative FFT in 10th chirp, the estimated thermal noise floor was found to be $\approx$ -16.1~dB, while the noise floor increased to about 4.8~dB due to the presence of interference in the 50th chirp. Thus, the interference power $Threshold$ was set as -6~dB. It is noted that the thermal noise floor in the negative half of FFT does not change as it is determined by the receiver itself, and there are no targets in the negative half of FFT as we previously discussed in Sec. \ref{3b1}, therefore the interference power threshold setting can be fixed and only varies for different hardware. 

\par As stated in Algorithm \ref{alg1}, because the noise power in the negative FFT was less than the $Threshold$, the ANC was bypassed for the 10th chirp. On the other hand, the ANC with filter length $L$ of 8, and a step size $\Delta w$ of \(\frac{2}{30*P}\) was applied for the 50th chirp data. Consequently, the target was now clearly visible in Fig. \ref{Figure13} (g), with an increased SIR of 7.6~dB compared to Fig. \ref{Figure13} (d) under a CFAR window with 20 reference cells and 6 guard cells.

\begin{figure}[H]
	\centering
	\includegraphics[width=3in]{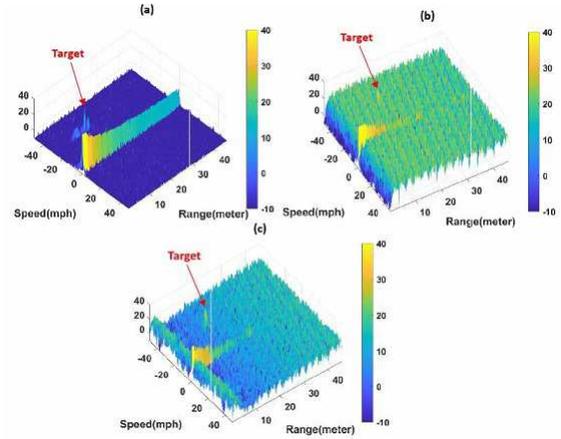}
	\caption{Data process of one CPI in the field experiment. (a) Range-Doppler of raw data without interference, SIR~=~32.82~dB. (b) Range-Doppler of raw data with interference, SIR~=~15~dB. (c) The proposed method output of raw data in (b), SIR~=~28.2~dB.}
	\label{Figure14}
\end{figure}
\par Furthermore, the Doppler-FFT was applied on the filtered range-FFT data. Fig. \ref{Figure14} (c) showed the final range-Doppler map, with 13.2~dB increased SIR (under CFAR window of 20 reference cells and 6 guard cells in both the range and Doppler dimension) compared to Fig. \ref{Figure14} (b) without the proposed method. However, the close range interference was not effectively mitigated in Fig. \ref{Figure14} (c), as the adaptive filter was still adapting to converge to the optimum solution in the beginning. As a comparison, Fig. \ref{Figure14} (a) showed the range-Doppler map of a similar target when the aggressor radar was off.

\section{Conclusion}
To address the increased noise floor due to the interference from the aggressor radar, we proposed a novel interference mitigation method based on adaptive noise canceller, which takes the positive and negative half of FFT as the input of its primary and reference channel, respectively. In a $Matlab$ simulation, this method showed a very good SIR improvement about 6~dB with a proper adaptive filtering step size. As we found, the step size is a trade-off between the increased SIR and side lobes. And in an experiment for a moving vehicle at about 15 meters, this method could achieve 7.6~dB SIR increment in range-FFT data and 13.2~dB in range-Doppler map albeit the performance was limited in the very close range.


%

%
%
%
%
%

\ifCLASSOPTIONcaptionsoff
  \newpage
\fi



\bibliographystyle{IEEEtran}
\bibliography{IEEEabrv,./Automotive_Radar_Interference_Mitigation_using_Adaptive_Noise_Canceller}
%

%

\vskip 0pt plus -1fil

\begin{IEEEbiography}[]{Feng Jin}
(S'19) received the B.S. and M.S. degrees in electrical engineering from Beihang University, Beijing, China, in 2011 and 2014, respectively. He is currently working toward the Ph.D. degree at University of Arizona, Tucson, Arizona, USA. His research interests include automotive radar signal processing and machine learning on mmWave radar sensor for object classification.
\end{IEEEbiography}

\vskip -2pt plus -1fil

\begin{IEEEbiography}[]{Siyang Cao}
(S'11-M'15) received the Ph.D. degrees in Electrical Engineering from The Ohio State University in Columbus, Ohio in 2014. He received the B.S. degree in electronic and information engineering from Xidian University in Shanxi, China in 2007, and the M.S. degree in circuits and system from South China University of Technology in Guangdong, China in 2010. Since August 2015, he has been an Assistant Professor in the Electrical and Computer Engineering Department of The University of Arizona in Tucson, AZ, USA. His research interests include radar waveform design, synthetic aperture radar, commercial radar, and signal processing with emphasis on radar signal.
\end{IEEEbiography}





\end{document}